\documentclass[fleqn,usenatbib]{mnras}

\usepackage[T1]{fontenc}
\usepackage{ae,aecompl}

\usepackage{graphicx}	
\usepackage{amsmath}	
\usepackage{amssymb}	
\usepackage{xcolor}

\def\be{\begin{equation}}
\def\ee{\end{equation}}

\newcommand\quotes[1]{``{#1}"}
\def\gsim{\lower.5ex\hbox{\gtsima}} 
\def\lsim{\lower.5ex\hbox{\ltsima}} 
\def\gtsima{$\; \buildrel > \over \sim \;$} 
\def\ltsima{$\; \buildrel < \over \sim \;$} \def\gsim{\lower.5ex\hbox{\gtsima}} 
\def\lsim{\lower.5ex\hbox{\ltsima}} 
\def\simgt{\lower.5ex\hbox{\gtsima}} 
\def\simlt{\lower.5ex\hbox{\ltsima}}

\def\mum{\mu {\rm m}}

\def\Sg{$\Sigma_{\rm gas}$}
\def\S*{$\Sigma_{\rm SFR}$}
\def\Scii{$\Sigma_{\rm [CII]}$}
\def\Sciii{$\Sigma_{\rm CIII]}$}

\def\HI{\hbox{H~$\scriptstyle\rm I\ $}}

\def\CIIion{\hbox{C~$\scriptstyle\rm II $}}

\def\ks{\kappa_{\rm s}}

\definecolor{apcolor}{HTML}{b3003b}
\definecolor{afcolor}{HTML}{800080}
\definecolor{lvcolor}{HTML}{0066cc}
\definecolor{mdcolor}{HTML}{01abdf} 
\definecolor{cbcolor}{HTML}{ff0000}
\definecolor{sccolor}{HTML}{cc5500} 
\definecolor{sgcolor}{HTML}{00cc7a}

\defcitealias{carniani2018}{C18}
\defcitealias{ferrara2019}{F19}

\title[SF law in the EoR]{Star formation law in the EoR from [CII] and CIII] lines}

\author[Vallini et al.]{
L. Vallini$^{1}$\thanks{E-mail: vallini@strw.leidenuniv.nl (LV)},
A. Ferrara$^{2}$,
A. Pallottini$^{2,3}$,
S. Carniani$^{2}$,
S. Gallerani$^{2}$
\\
$^{1}$Leiden Observatory, Leiden University, PO Box 9500, 2300 RA Leiden, The Netherlands\\
$^{2}$Scuola Normale Superiore, Piazza dei Cavalieri 7, 56126 Pisa, Italy\\
$^{3}$Centro Fermi, Museo Storico della Fisica e Centro Studi e Ricerche \quotes{Enrico Fermi}, Piazza del Viminale 1, Roma, 00184, Italy\\
}

\date{Accepted XXX. Received YYY; in original form ZZZ}

\pubyear{2018}

\begin{document}
\label{firstpage}
\pagerange{\pageref{firstpage}--\pageref{lastpage}}
\maketitle

\begin{abstract}
We present a novel method to simultaneously characterize the star formation law and the interstellar medium properties of galaxies in the Epoch of Reionization (EoR) through the combination of [CII]158$\mu$m (and its known relation with star formation rate) and CIII]$\lambda$1909\AA~emission line data. The method, based on a Markov Chain Monte Carlo algorithm, allows to determine the target galaxy average density, $n$, gas metallicity, $Z$, and \quotes{burstiness} parameter, $\kappa_s$, quantifying deviations from the Kennicutt-Schmidt relation. As an application, we consider COS-3018 ($z=6.854$), the only EoR Lyman Break Galaxy so far detected  in both [CII] and CIII]. We show that COS-3018 is a moderate starburst ($\ks \approx 3$), with  $Z \approx 0.4 \, Z_{\odot}$, and  $n \approx 500\, {\rm cm^{-3}}$. Our method will be optimally applied to joint ALMA and JWST targets.  
\end{abstract}
\begin{keywords}
galaxies: ISM -- galaxies: high-redshift -- ISM: photodissociation region
\end{keywords}
\section{Introduction}
How do galaxies convert their gas into stars? How do their interstellar medium (ISM) properties influence star formation?
Answers to these questions hold the key to understand galaxy evolution \citep[][for a review]{dayal2018}.

In nearby galaxies and at intermediate redshifts, the so-called Kennicutt-Schmidt (KS) law \citep{schmidt1959, kennicutt1998, delosreyes2019}, relating the star formation rate and the gas surface density is well established. Dozens of observational studies ranging from the local Universe \citep[e.g.][]{bigiel2008, schruba2011} up to $z\approx 3-4$ \citep[e.g.][]{daddi2010, tacconi2013, genzel2015, hodge2015} have shown that, when averaged over kpc-scales, the star formation rate, \S*, and the cold gas, \Sg, surface density in disk galaxies follow a tight relation,
\begin{equation}
\Sigma_{\rm SFR}=10^{-12}\kappa_s \Sigma^{m}_{\rm g}, \qquad (m\approx 1.4)
\label{KS}
\end{equation}
valid over about five dex in \Sg~\citep{heiderman10}. 
Two different SF regimes can be identified in the \S*$-$\Sg~plane: \quotes{quiescent} ($\kappa_s \approx 1$) and \quotes{starburst} ($\kappa_s>1$) galaxies \citep{daddi2010, hodge2015}.
However, as we move towards the Epoch of Reionization (EoR, $z>6$), a precise assessment of the KS relation becomes progressively more difficult or even impossible. Advanced optical/near-infrared facilities such as the Hubble Space Telescope (HST), Very Large Telescope (VLT), Keck, and Subaru telescopes enabled rest-frame ultraviolet (UV) continuum and line emission detection in large samples of EoR galaxies \citep[e.g.][]{bouwens2015}. The exquisite spatial resolution of such instruments often allows to carry out UV size measurements  \citep[e.g.][]{shibuya2015, curtis-lake2016, bowler2017, kawamata2018, matthee2019}, hence enabling estimates of \S* in EoR galaxies.

Spatially-resolved detections of cold gas tracers such as CO lines in these systems are instead still very challenging or lacking \citep[e.g.][]{vallini2018, dodorico2018, pavesi2019}. 
Low-$J$ ($J\leq3$) CO rotational transitions detections in quiescent star-forming galaxies are still limited to $z\simlt3$ \citep{tacconi2013, genzel2015} while at $3<z<4.5$ only few significant detections have been reported, either in massive sub-millimeter galaxies \citep{hodge2015, sharda2018, sharda2019} or  strongly lensed systems \citep[]{coppin2007, dessauges-zavadsky2015}. Beside CO, alternative tracers such as the [CII]158$\mu$m emission, calibrated at $z\approx 1$, have been proposed to measure the gas mass \citep{zanella2018}. However, \citet{pavesi2019} pointed out that a full characterization of the ISM in EoR galaxies requires, in addition to [CII], information on CO or a ionized gas tracer.

In the last years ALMA \citep[][]{carilli2013} has opened a new window on the ISM properties of early galaxies, allowing for the first time the detection at high spatial resolution and sensitivities 
of the $158\mum$ $^{2}P_{3/2}\rightarrow \,^{2}P_{1/2}$ transition of ionized carbon ([CII]) \citep[e.g.][]{maiolino2015, capak2015, smit2018, carniani2018, hashimoto2019, matthee2019}. [CII] is the most luminous line in the far-infrared (FIR) band \citep{hollenbach1999}, and traces cold neutral/molecular gas associated with Photo Dissociation Regions (PDR) \citep[e.g.][]{vallini2015, pallottini2017, ferrara2019}. 

Importantly, the James Webb Space Telescope (JWST) will soon provide a complementary probe of the high-$z$ ISM, targeting rest-frame optical/UV emission lines associated with ionized gas. Among the various UV line tracers, current state-of-the-art observational campaigns with e.g. VLT and KECK \citep[e.g.][]{stark2015, stark2017, ding2017, laporte2017, mainali2018, hutchison2019} and theoretical studies \citep[][]{feltre2016,jaskot2016, nakajima2017} showed that low-metallicity, $z>5$ galaxies are expected to show prominent C III]$\lambda$1909\AA~ line emission. Such line will be likely detected in large samples of galaxies with JWST, opening an interesting synergy with ALMA in targeting carbon lines.
In this work we show that, by combining [CII]158$\mu$m and CIII]$\lambda 1909$\AA~ data it is possible to constrain \emph{at the same time} the KS relation and ISM properties of EoR galaxies. After presenting the method in Sec. \ref{sec:method}, in Sec. \ref{sec:results} we apply it to COS-3018555981 at $z=6.854$, the only Lyman Break Galaxy (LBG) representative of quiescent star-forming galaxies so far detected at $z>6.5$ in [CII] and CIII]. In Sec. \ref{sec:conclusions} we discuss the implications of the results and our conclusions. 

\section{Method}\label{sec:method}
Our method is based on an extension of the physical model for the [CII] emission in galaxies presented in  \citet[][F19 hereafter]{ferrara2019}. While locally a tight \Scii$-$\S* correlation has been measured \citep{delooze2014, herrera-camus2015}, many EoR galaxies show \Scii~values almost systematically fainter than expected from their measured \S* \citep{carniani2018, pallottini:2019}.

\citetalias{ferrara2019} argued that three factors can produce such deficit: (a) a high \quotes{burstiness} parameter $\ks$ (see eq. \ref{KS});  
(b) a low gas density $n$; (c) a low gas metallicity $Z$, with (b) and (c) playing a sub-dominant role. If a third observed quantity, beside \Scii~and the deviation, $\Delta_{\rm [CII]}$, from the local \Scii$-$\S* relation  is available, then the $n$, $Z$, $\ks$ degeneracy can be broken, thus enabling a complete characterization of the ISM properties and star formation law in EoR galaxies. 

CIII] is an excellent additional candidate line to break the degeneracy. In fact, contrary to the [CII] line, its luminosity at fixed \S* grows with $\kappa_s$ due to the progressively thicker ionized layer (\citetalias{ferrara2019}). Moreover, the \Sciii/\Scii~ratio is unaffected by the (unknown) relative abundances of different elements at high-$z$. In what follows we summarize the basic equations of our model, and operationally define $\Delta_{\rm [CII]}$. We refer the interested reader to \citetalias{ferrara2019} for a complete derivation of the equations.

\subsection{[CII] emission model}
\label{sec:cii_model}
Consider a disk galaxy with mean gas density $n$, carbon abundance ${\cal A_C}=2.7\times 10^{-4}$  \citep{asplund2009}, metallicity $Z$, and ionization parameter $U = n_\gamma/n$. The [CII] surface brightness [$\rm  L_\odot\, kpc^{-2}$], can be written (\citetalias{ferrara2019}, eq. 35) as:
\begin{equation}
\Sigma_{\rm [CII]} = 2.4 \times 10^9 \, F_{\rm [CII]} (n, Z, U)
\label{eq:scii}
\end{equation}
where $F_{\rm [CII]} =f_{\rm [CII]}^{i} + f_{\rm [CII]}^{n}$ (in $\rm erg \, s^{-1}\, cm^{-2}$) is the emerging [CII] flux.
The first term in the previous equation accounts for the emission due to collision of C$^+$ ions with $e^-$ in the ionised layer: 
\begin{equation}
f_{\rm [CII]}^{i} = n_e \Lambda^{(4)}_e {Z} {\cal A}_{C}  N_{\rm HI}(Z,U),
\label{eq:cii_ion}
\end{equation}
where $n_e\approx n$ is the number density of free electrons, $\Lambda^{(4)}_e(T=10^{4}{\rm K}) =  1.2 \times 10^{-21}$ erg cm$^3 \rm s^{-1}$ is the cooling rate (Appendix B, \citetalias{ferrara2019}), and $N_{\rm CII}\approx {\cal A}_{C} Z N_{\rm HI}$ is the C$^+$ column density in the ionized layer. The second term, $f_{\rm [CII]}^{n}$, accounts for the emission due to collisions with H atoms in the neutral ($T=10^2\, \rm K$) part of the Photo Dissociation Region (PDR):
\begin{equation}
f_{\rm [CII]}^{n} = n \Lambda^{(2)}_H {\cal A}_{C} {Z} N_{\rm PDR}(Z,U).
\label{eq:cii_neut}
\end{equation}
In the previous equation, $n$ is the \HI number density; $\Lambda^{(2)}_H = 7.65 \times 10^{-24}$ erg cm$^3 \rm s^{-1}$ is the collisional cooling rate (Appendix B, \citetalias{ferrara2019}), and the \CIIion\, column density is $N_{\rm CII}\approx {\cal A}_C Z N_{\rm PDR}$.
In Eq.s \ref{eq:cii_ion}-\ref{eq:cii_neut} $N_{\rm HI}$ and $N_{\rm PDR}$ depend on the dust shielding of the intensity (parametrized by $U$) of the ionizing interstellar radiation field. We assume  a constant dust-to-gas ratio, so that the dust column density providing the extinction is $\propto Z$. Rewrite eq. 15 of \citetalias{ferrara2019} in terms of these two quantities:
\begin{equation}
N_{\rm HI} = 3.7\times 10^{17} {\rm ln}\left(1-\frac{1+59 Z\, U}{1+21.7 Z\, U}\right)\,;
\end{equation}
also use eq. 30 of \citetalias{ferrara2019} to write
\begin{equation}
N_{\rm PDR} =  {\rm min} \left[ \frac{1.7\times 10^{21}}{Z} {\rm ln}\left( 1+\frac{10^5 U}{1+0.9 Z^{1/2}}\right), \, N_0 \right] - N_i \,,
\end{equation}
where $N_0$ is the disk total gas column density, and 
\begin{equation}\label{eq:colum_ion}
N_i = 1.7\times 10^{21} Z^{-1} {\rm ln} \frac{1+59 ZU}{1+21.7 ZU}
\end{equation}
is the ionized layer column density (eq. 14, \citetalias{ferrara2019}). 

Finally, $U$ can be related to the gas surface density, \Sg$=7.5\times10^7 N_{0,22}\, \rm M_{\odot} \,kpc^{-2}$ (\citetalias{ferrara2019}), and to \S*~as
\be
U = 1.7\times 10^{14}\,\frac{\Sigma_{\rm SFR}}{\Sigma_g^2} \simeq 10^{-3} \, \ks^{10/7} \Sigma_{\rm SFR}^{-3/7}.
\label{Udef}
\ee
where we substituted the KS relation (eq. \ref{KS}) to extract the dependence on $\kappa_s$. From Eq. \ref{eq:scii} we then predict\footnote{The [CII] flux obtained with the above analytical model has been shown to be in excellent agreement with CLOUDY \citep{ferland2017} full RT calculations (Fig. 3, \citetalias{ferrara2019}).} the \Scii$-$\S*~relation for a given set  ($n$, $Z$, $\ks$):
\begin{equation}
\label{eq:cii_modello}
\Sigma_{\rm [CII]} = 2.4 \times 10^9 \, F_{\rm [CII]} ( \Sigma_{\rm SFR}|n, Z, \ks).
\end{equation}

\subsection{CIII] emission model.}
Following the same reasoning outlined in Sec \ref{sec:cii_model}, the CIII] surface brightness can be written as:
\begin{equation}
\Sigma_{\rm CIII]} = 2.4 \times 10^9 \, F_{\rm CIII]} (n, Z, U).
\label{eq:sciii}
\end{equation}
The CIII] emission is produced by collisional excitation of $\rm C^{2+}$ ions by free electrons. Hence,
\begin{equation}
F_{\rm CIII]} = n_e \Lambda_{\rm CIII]} {\cal A}_{C} {Z}  N_i(Z, U),
\label{eq:fciii}
\end{equation}
where the cooling rate at $T=10^4$ K $\Lambda^{(4)}_{\rm CIII]} = 5.8 \times 10^{-22}$ erg cm$^3 \rm s^{-1}$ (Appendix B, \citetalias{ferrara2019}), the $\rm C^{2+}$ column density is $N_{\rm CIII}\simeq {\cal A}_{C} {Z}  N_i$, with $N_i$ as in in Eq. \ref{eq:colum_ion}. Use Eq. \ref{eq:sciii} to predict the \Sciii$-$\S*~relation for a given set  ($n$, $Z$, $\ks$):
\begin{equation}\label{eq:ciii_modello}
\Sigma_{\rm CIII]} = 2.4 \times 10^9 \, F_{\rm CIII]} (\Sigma_{\rm SFR}| n, Z, \kappa_s).
\end{equation}

\subsection{Deviations from the local \texorpdfstring{$\Sigma_{\rm [CII]}-\Sigma_{*}$}{} relation}
In the local Universe a well-assessed \Scii$-$\S*~relation is found in spiral \citep{herrera-camus2015} and low-metallicity dwarf galaxies \citep{delooze2014}. In the last few years, the extension of such relation to EoR galaxies has become feasible thanks to the high spatial resolution of ALMA observations. As noted by e.g., \citet{carniani2018} and \citetalias{ferrara2019}, most of the sources at $z>5$ are found to have \Scii \,values fainter than expected on the basis of the local relation.   
In what follows, we will adopt the following functional form of the \Scii $-$ \S* relation,
\be
0.93 \log\Sigma^{local}_{\rm [CII]} = \log \Sigma_{\rm SFR} + 6.99 \,,
\label{dLfit}
\ee
which has a small 1$\sigma$ dispersion of 0.32 dex. This is obtained by \citet{delooze2014} for low-metallicity dwarf galaxies. We checked that using a different relation \citep[e.g.][]{herrera-camus2015} does not affect our results. These systems are usually considered to be fair analogs of reionization sources.  We define the expected deviation from the local \Scii relation (at fixed \S*) as 
\begin{equation}
\label{eq:deviazione_modello}
 \Delta_{\rm [CII]} (\Sigma_{\rm SFR}| n, Z, \kappa_s) \equiv \log\,\Sigma_{\rm [CII]} - \log\,\Sigma^{local}_{\rm [CII]}.
\end{equation}
Note that is $\Delta_{\rm [CII]}$ is a function of the three parameters ($\ks$, $n$, $Z$).
As a caveat, it is worth stressing that: \emph{(i)} in some cases the complex morphology of early galaxies \citep[e.g.][]{kohandel2019} makes the determination of the actual size of the [CII] emitting region somewhat challenging; \emph{(ii)} in the local \Scii-\S*~calibration, the SFR is obtained both by optical lines (e.g. H$\alpha$) and the FIR continuum. On the contrary, the \S*~in EoR galaxies is often derived from the UV continuum only, as the majority of high-$z$ sources are undetected in dust continuum, and those detected have only one point on the dust continuum SED. This makes the determination of the total infrared luminosity highly dependent on the unknown dust temperature \citep[][]{behrens2018}. 

\subsection{Parameters derivation}
Eq. \ref{eq:cii_modello}, \ref{eq:ciii_modello}, and \ref{eq:deviazione_modello} allow us to solve for the three unknown parameters ($\ks$, $n$, $Z$). Our solution method is based on a Bayesian Markov Chain Monte Carlo (MCMC) framework. We use the $\chi^2$ likelihood function to fit the observed \Scii, \Sciii,\, and $\Delta_{\rm [CII]}$ of a galaxy and determine the posterior probability distribution of the model parameters. This choice enables us to fully characterise any potential degeneracies between our model parameters, while also providing the individual probability distribution functions (PDFs) for each of them. In this work we use the open-source \texttt{emcee} Python implementation \citep{foreman2013} of the Goodman \& Weare's Affine Invariant MCMC Ensemble sampler \citep{goodman2010}. 
\begin{table}
\centering
\caption{Observed properties of COS-3018. Data from \citet{carniani2018} (1), \citet{smit2018} (2), \citet{laporte2017} (3).}
\medskip
\begin{tabular}{llc}
\hline
Quantity & Value & Reference \\
\hline
$r_{\rm UV}$ (kpc) & $1.3\pm0.1$ & (1) \\
SFR$_{\rm UV}$ ($\rm M_{\odot}\, yr^{-1}$) & $18.9\pm 1.5$ & (1) \\
\S*~($\rm M_{\odot} \, yr^{-1} kpc^{-2}$) & $3.6\pm 0.5$ & (1)\\
$L_{\rm [CII]}$ ($10^8\, L_{\odot}$) & $4.7\pm 0.5$ & (2) \\
$r_{\rm [CII]}$ (kpc) & $2.6\pm 0.5$ & (1) \\
\Scii~($L_{\odot} \, \rm kpc^{-2}$) & $(2.2\pm 0.7) \times 10^{7}$ & This work\\
$L_{\rm CIII]}$ ($L_{\odot}$) & $(1.9\pm0.4)\times 10^{8}$ & (3)\\
\Sciii~($L_{\odot} \, \rm kpc^{-2}$) & $(3.7\pm 0.4) \times 10^{7}$& This work \\
\hline
\end{tabular}
\end{table}
\section{A case study: COS-3018}
\label{sec:results}
As a case study, we apply our model to COS-3018555981 (COS-3018 herafter), the only Lyman Break Galaxy \citep{smit2018} in the EoR ($z \approx 6.85$) so far detected both in [CII]
and CIII] \citep{smit2018, laporte2017}. \footnote{There is a (lensed) source (A383-5.1/5.2) at slightly lower redshift \citep[$z=6.02$,][]{richard2011}, detected both in [CII] and CIII] \citep{knudsen2016, stark2015}. However, given that (i) [CII] and CIII] emission have been observed in two different images thus requiring a careful evaluation of the impact of the lens model, and (ii) the [CII] emission is only tentatively resolved, hence yielding uncertainties in the estimate of \Scii, we defer a detailed study to a forthcoming paper.}
\begin{figure*}
    \centering
    \includegraphics[scale=0.6]{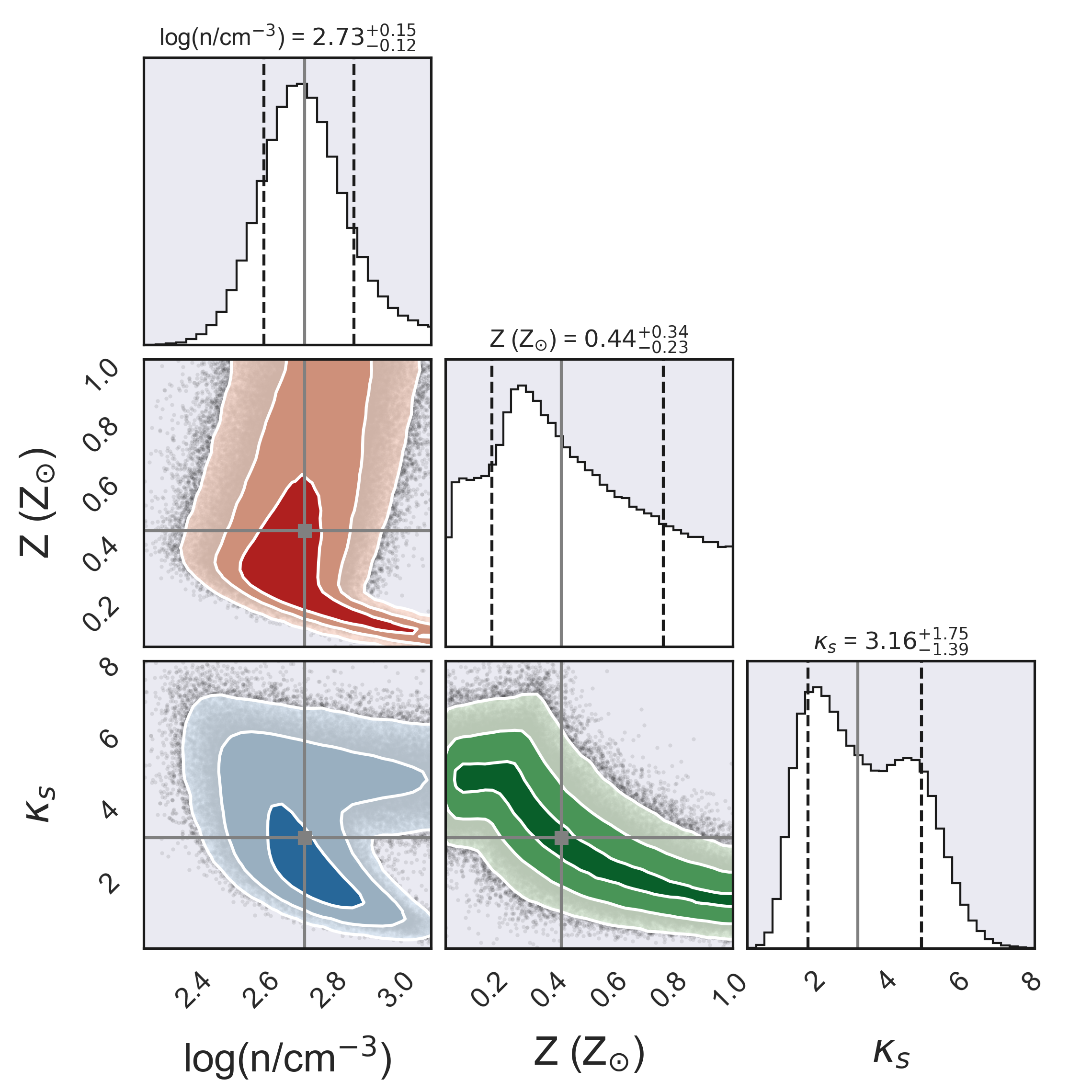}
    \caption{Corner plot showing the posterior probability distributions of $\log n$, $Z$, and $\kappa_s$ for COS-3018 at $z=6.854$. The contours represent 1$\sigma$, 2$\sigma$, and 3$\sigma$ levels for the 2D distributions. The best-fit parameters and the 16\%, 84\% percentiles are plotted with grey squares and dashed lines, respectively. \label{fig:mcmc_plot}}
\end{figure*}

COS-3018 was first discovered by \citet{tilvi2013, bowler2014}, and then re-analyzed by \citet{smit2015} as a part of their selection of IRAC excess sources in the 3.6 or 4.5 $\mu$m photometric bands deriving $z_{phot}=6.76$.
\citet{smit2018} spectroscopically confirmed the source at $z=6.854$ via the detection of the [CII] 158$\mu$m line. 
The [CII] lu\-mi\-no\-si\-ty of COS-3018 is $L_{\rm [CII]}=(4.7\pm 0.5)\times10^{8}\,\rm L_{\odot}$ \citep{smit2018}.
Both the spatially resolved UV and [CII] emission have been re-analyzed by \citet{carniani2018} who, by using the \citet{kennicutt2012} UV-star formation rate (SFR) calibration ($\log ({\rm SFR}/{\rm M_{\odot}\,yr^{-1}})=\log({\rm L_{UV}}/{\rm erg\,s^{-1}})- 43.35$) derived ${\rm SFR}_{\rm UV}=18.9\pm 1.5\, \rm M_{\odot}\, yr^{-1}$. The \citet{kennicutt2012} relation assumes Kroupa initial mass function, $\approx 10 \, \rm Myr$ as mean stellar age producing the UV emission, and $Z=Z_{\odot}$ for the stellar metallicity. The galaxy is instead undetected in dust continuum \citep{smit2018}.
COS-3018 is a compact galaxy: the size of the [CII] emitting region is $r_{\rm [CII]}=2.6 \pm 0.5$ kpc \citep{carniani2018}, while the star forming region traced by the rest-frame UV emission is considerably smaller, $r_{\rm UV}=1.3 \pm 0.1$ kpc \citep{carniani2018}. As both [CII] and UV emissions are marginally resolved, we can compute the [CII] surface brightness \Scii$=L_{\rm [CII]}/\pi r^2_{\rm [CII]} = (2.2\pm 0.7) \times 10^{7}\,\rm L_{\odot} \, \rm kpc^{-2}$, and the SFR surface density \S*$={\rm SFR_{\rm UV}}/\pi r^2_{\rm UV} =3.6 \pm 0.5 \, \rm M_{\odot}\, yr^{-1}\, kpc^{-2}$. This translates into $\Delta_{\rm [CII]} = -0.74$.

The CIII]$\lambda$1909\AA~emission has been detected with XSHOOTER/VLT at 4$\sigma$ \citep[$f_{\rm CIII]}=1.33 \pm 0.31 \times 10^{-18}\,{\rm erg}\, {\rm s}\, {\rm cm}^{-2}$][]{laporte2017}, yielding $L_{\rm CIII]} = (1.9 \pm 0.4) \times 10^{8} \rm \, L_{\odot}$ and, by assuming $r_{\rm UV}$ to be a proxy of the size of the nebular line emitting region, we derive \Sciii$=L_{\rm CIII]}/\pi r^2_{\rm UV} =(3.7 \pm 0.4) \times 10^7 \, \rm L_{\odot} \, kpc^{-2}$. This is a reasonable assumption as CIII] and UV continuum trace ionized gas, and the CIII] 1D spectrum is extracted from the UV emitting region. All quantities are presented in Tab. 1.

From the MCMC procedure we derive the best-fit ($\ks, n, Z$) values and confidence intervals for COS-3018. 
We run \texttt{emcee} with $100$ random walkers exploring the parameter space for $5 \times 10^4$ chain steps.
The chains have been initialised by distributing the walkers in a small region around $Z=0.2\rm\, Z_{\odot}$, $\log (n/{\rm cm^{-3}}) =2.5$ and $\kappa_s =1$. The $Z=0.2\rm \, Z_{\odot}$ value is equal to that assumed by \citet{bowler2014} for the stellar metallicity in their SED fitting of COS-3018. As a caveat we note that differences between stellar and gas metallicities are likely to occur in high-$z$ systems \citep{steidel2016}.

We assume uniform priors for the gas density in the range $1.0 \leq \log (n/{\rm cm^{-3}}) < 3.3$, metallicity $0.05 < (Z/Z_{\odot}) \leq 1.0$, and burstiness parameter $0.1 \leq \ks \leq 50$. To estimate the effective number of independent samples we calculate the $n_{burn}$ steps necessary to ensure chain independence. We adopt $n_{burn}= 50 \tau$, where $\tau=147$ is the average auto-correlation time in chain steps computed with the built-in implementation given in \texttt{emcee}\footnote{The sampler should be run for $>10 \tau$ steps before walkers fill the relevant parts of parameter space and becomes an independent set of samples from the distribution \citep{foreman2013}.}. Next, we discard the burn-in chunk and use the remaining portion to sample the posterior probability. The mean acceptance fraction is 0.504. For the best fit parameters and confidence levels we use the median and the 16th, 84th percentiles of the marginal PDFs \citep{foreman2013}. 

The result of the MCMC analysis is shown in Fig. \ref{fig:mcmc_plot}. The best fit gas density in COS-3018 is $\log (n/{\rm cm^{-3}}) = 2.73^{+0.15}_{-0.12}$ which is in an excellent agreement with the mean gas density $\log (n/{\rm cm^{-3}}) \approx 2.5$ of dense neutral/molecular gas  found by cosmological zoom-in simulations of prototypical LBGs at $z\approx 6-7$ \citep{pallottini:2019}.
Such high density value might also explain the non-detection of the CIII]$\lambda$1907\AA~line \citep{laporte2017} in this object. 

The best fit burstiness parameter is $\ks = 3.16^{+1.75}_{-1.39}$, implying that COS-3018 is a moderate starburst galaxy. Note that the presence of an ongoing starburst in COS-3018 has been tentatively suggested by previous studies \citep{smit2018} because of the high equivalent-width of optical emission lines (EW([O III]+H$\beta$)=$1424\pm 43$\AA) \citep{smit2015}. The $\kappa_s$ distribution resulting from our analysis shows a double-peak profile, with a  lower peak at $\kappa_s \approx 2$ and the higher one at $\kappa_s \approx 5$.
We explain this behaviour as follows. 
There are two possibilities to reproduce the observed [CII]/CIII] ratio. The first one corresponds to a low-metallicity solution with $Z\approx 0.2 \, \rm Z_{\odot}$, and $\kappa_s \approx 5$. In this case, the [CII] flux is higher as $N_{\rm PDR} \propto Z^{-1}$. 
To compensate for the [CII] increase, higher $\ks$, and consequently higher $U$, values are required, resulting in large ionized gas column densities boosting the CIII] emission (see \citetalias{ferrara2019}). The second peak at $\kappa_s \approx 2$ corresponds to a higher metallicity ($Z\approx 0.7\rm \, Z_{\odot}$) which produces a thinner PDR region emitting the [CII]. In this situation, in order to fit the observed ratio a lower $\ks$ values is obtained. Note that the plateau at $Z \simlt 0.2\, \rm Z_{\odot}$ happens because in this regime the [CII] luminosity is independent on $Z$ as $N_{\rm HI} > N_0$  and hence $N_{\rm PDR}\approx N_0$ (see \citetalias{ferrara2019}). In this region of the parameter space, $Z$ is essentially unconstrained. 
Finally, our MCMC analysis constrains the gas-phase metallicity of COS-3018 in the range $Z = 0.44^{+0.34}_{-0.23} \rm \, Z_{\odot}$ that, despite the large scatter due to the above considerations, allows us to safely conclude that COS-3018 is less chemically evolved than the Milky Way but not extremely metal poor. 

Note that while our method simultaneously constrains $\kappa_s$, $Z$ and $n$, in principle one can use the [CII] luminosity alone to estimate the burstiness parameter. Extrapolating the $z\approx 1$ relation $M_{g}= 30 L_{\rm [CII]}$ \citep[mean absolute deviation of 0.2 dex,][]{zanella2018} to the EoR, one finds $\Sigma_{g} = 30 L_{\rm [CII]}/\pi r^2_{\rm [CII]} = (7.2 \pm  3.6)\times 10^8 \, \rm M_{\odot} \, kpc^{-2} $. By inverting Eq. \ref{KS}, we get $\kappa_s = 1.43 \pm 1.0$ which is consistent within $\approx 1\sigma$ with the best fit $\kappa_s$ found with our MCMC. 

\section{Discussion and conclusions}
\label{sec:conclusions}
We have presented a novel method to simultaneously determine  the star formation law, gas density and metallicity of galaxies in the EoR. This is done by exploiting the [CII] and CIII] surface brightness, and the deviation from the local \Scii $-$ \S* relation. The method is based on a MCMC algorithm that allows us to determine the best fit $\kappa_s$ , $n$, and $Z$ of a galaxy, and their confidence levels. In particular, we analyzed the case of COS-3018 a LBG at $z=6.854$, finding that it is a moderate starburst galaxy ($\ks = 3.16^{+1.75}_{-1.39}$), with sub-solar gas-phase metallicity ($Z = 0.44^{+0.34}_{-0.23} \rm \, Z_{\odot}$) and a mean gas density of $\log (n/{\rm cm^{-3}}) = 2.73^{+0.15}_{-0.12}$, in very nice agreement with predictions from state-of-the-art simulations of EoR galaxies \citep{pallottini:2019}.

The only other LBG at the end of the EoR for which the KS relation has been constrained is HZ10, for which \citet{pavesi2019} estimated $\Sigma_g \approx 10^{10} \, \rm{M_{\odot}\, kpc^{-2}}$ and $\Sigma_* \approx 10^{1.2} \, \rm{M_{\odot}\, yr^{-1}\, kpc^{-2}}$. They also point out that HZ10 has a very low $\kappa_s\approx 0.1$ value, compared to the other LBG in their sample (HZ6) which was undetected in CO. The HZ10/HZ6 CO luminosity ratio is $> 6.5$, in spite of a more modest factor of 3 in their SFR ratio. They proposed that the difference in CO luminosity could be due to: (1) variation in star formation efficiency and/or (2) low-$Z$/dust abundance suppressing CO emission in HZ6. Our method can help clarifying this point in these and similar EoR systems.
In spite of the success of the method, there are some caveats to keep in mind. The first one is that, by construction, the ISM of the galaxy is approximated with a single gas slab, with an unique density and $Z$. This is obviously a simplification as the [CII] and CIII] emission might not be fully co-spatial. Moreover, different gas phases in the ISM show density variations throughout the galaxy.

Nevertheless, our method offers the first glimpse of global (spatially averaged) properties of EoR galaxies. As essential\-ly no alternative constraints are currently available for EoR sources, our model can provide a first order estima\-te of their key ISM properties. The obvious advantage is that it can constrain metallicity, mean gas density and - more importantly - the SF law in large samples of sources by using only two emission lines that are detectable by ALMA, current optical/NIR telescopes and, in the near future, JWST. 

\section*{Acknowledgements}
LV is supported by a Marie Sk\l{}odowska-Curie fellowship (grant agreement No. 746119). 
AF and SC are supported by the ERC-Adg INTERSTELLAR H2020/740120.
%
%
AF is partially supported by the C.F. von Siemens-Forschungspreis der Alexander von Humboldt-Stiftung Research Award.
\bibliographystyle{mnras}
\bibliography{master_biblio} 
\bsp	
\label{lastpage}
\end{document}